\documentclass[editedvolume,numreferences]{crckbked}
\usepackage{graphics}
\usepackage{graphicx}
\usepackage{epsfig}
\begin{document}

\begin{article}

\begin{opening}
\title{Few-electron liquid and solid phases in
artificial molecules at high magnetic field}
\runningtitle{Few-electron phases in artificial molecules}
\author{Massimo \surname{Rontani}\email{rontani@unimore.it}}
\author{Guido \surname{Goldoni}\email{goldoni@unimore.it}}
\author{Elisa \surname{Molinari}}
\runningauthor{Rontani, Goldoni, and Molinari}
\institute{INFM National Research Center on nanoStructures and \\
bioSystems at Surfaces (S$^3$) and Dipartimento di Fisica, \\
Universit\`a degli Studi di Modena e Reggio Emilia, \\
Via Campi 213/A, 41100 Modena, Italy}
\begin{abstract}
Coupled semiconductor quantum dots form artificial molecules where
relevant energy scales controlling the interacting ground state
can be easily tuned. By applying an external magnetic field it is
possible to drive the system from a weak to a strong correlation
regime where eventually electrons localize in space in an ordered
manner reminiscent of the two-dimensional Wigner crystal. We
explore the phase diagram of such ``Wigner molecules'' analyzing
the angular correlation function obtained by the Configuration
Interaction solution of the full interacting Hamiltonian. Focus is
on the role of tunneling in stabilizing different ground states.
\end{abstract}
\end{opening}

\section{Introduction}
\begin{figure}[htb]
\centerline{\epsfig{file=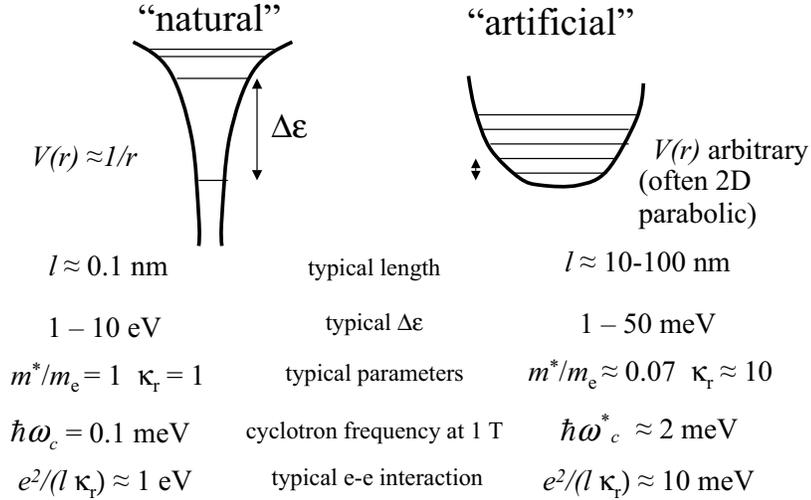,angle=90,width=11.6cm}}
\caption{Comparison between the typical parameters and energy scales
of ``natural'' and ``artificial'' atoms. $m^*$ and $\kappa_r$
are the effective mass and the static dielectric constant of the
host semiconductor, respectively. The (effective) cyclotron frequency
$\omega_c$ ($\omega_c^*$) is defined as $eB/mc$ ($eB/m^*c$), with
$B$ applied magnetic field.
}\label{fig1}
\end{figure}

Semiconductor quantum dots (QDs) are zero-dimensional systems
where a few electrons, typically $1<N<100$, are spatially confined
and the energy spectrum is completely discrete \cite{QDreview}.
Carriers can be injected one by one to the system in
single-electron transport \cite{PRLTarucha} or capacitance
\cite{Ashoori} experiments, based on the {\em Coulomb blockade}
\cite{Devoret} phenomenon, and the energy required to add one
electron can be measured if the electrostatic screening is poor
and the thermal smearing is low. Measurements of such ``electron
affinities'' \cite{PRLTarucha} have revealed a shell structure for
the correlated electron system and fine corrections to the energy
due to exchange interaction (Hund's rule): therefore QDs are often
regarded as {\em artificial atoms} \cite{MaksymPRL,Kastner}. In
Fig.~\ref{fig1} we compare the relevant parameters of natural and
artificial atoms: note that for the latter the typical energy
spacing is of the order of a few meV, less than the thermal energy
at experimentally reacheable temperatures \cite{QDreview}.
{\em Artificial molecules} can be built by coupling two QDs
together in a controlled way \cite{leoartificial,us}: this
additional degree of freedom enriches the physics of natural
molecules, since for the latter the inter-nuclear coupling is
almost fixed by the balance between nuclear repulsion and
electrostatic attraction mediated by valence electrons, while in
the former the stability of the electron system is externally
imposed.

Current focus on QDs stems from their technological
potentialities as optoelectronic devices (single-electron
transistors \cite{Devoret}, lasers \cite{laser}, micro-heaters and
micro-refrigerators based on thermoelectric effects
\cite{thermoelectric}) as well as from several proposals of
quantum information processing schemes in a solid-state
environment \cite{loss,filippo}. Here however we are more
interested in basic physical principles and we look at QDs as a
laboratory to explore the fundamentals of few-body strongly
interacting systems.

The plan of the paper is the following:
Section \ref{manipulating} is a primer on electron states in QDs. We
show which energy scales can be artificially tailored and their
effect on the interacting ground states. After, we specialize to the case of
the artificial molecule in very high magnetic field as a paradigma
of the strong correlation regime (Sec. \ref{Wigmol}).

\section{Manipulating the energy spectrum}\label{manipulating}
The critical issue of artificial atoms and molecules is that
almost all relevant energy scales can be controlled by means of
fabrication and/or tuning of experimental parameters such as
voltages or magnetic fields and made of comparable size.
\cite{QDreview,us,Reed}.
This is at difference with natural systems, where, for example,
the dominant energy scale is provided by the ionic potential
attracting the valence electrons, while the weaker
electron-electron interactions can be treated reasonably well by
mean field methods. Moreover, the orbital and spin coupling with
an external magnetic field is very small (cfr.~Fig.~\ref{fig1}).
The ground states are therefore determined by the successive
filling of the empty lowest-energy hydrogen-like orbitals,
according to the {\em Aufbau} theory. The open-shell
configurations are well described by Hund's rules, which can be
explained as the effect of the Coulomb term in the interacting
Hamiltonian.

In QDs the confining potential $V\!(\mbox{\boldmath $r$})$ is much
weaker, being provided by the electrostatic environment. Here
effective parameters, such as renormalized electron mass and
static dielectric constant of the host semiconductor determine the
kinetic energy, the strength of the Coulomb interaction, and the
coupling with the field (see Fig.~\ref{fig1}). The overall effect
is that there is no dominant energy scale (Fig.~\ref{fig1}),
making the problem of determining the electronic ground state much
more difficult.

In many of the experimental realizations of artificial atoms, the
system can be regarded as quasi two dimensional \cite{QDreview}:
since the confinement along the growth direction $z$ is much
tighter than in the $x-y$ plane, the $z$ degree of freedom is
frozen as far as low-energy states are concerned. In the effective
mass and envelope function approximation, the confinement potential
$V\!(\mbox{\boldmath $r$})$ can be decoupled as
\begin{equation}
V\!(\mbox{\boldmath $r$})=V(\mbox{\boldmath $\varrho$})+V(z),
\label{eq:decoupling}
\end{equation}
with $\mbox{\boldmath $\varrho$}\equiv (x,y)$. Lowest-energy
states are very well described by a two dimensional harmonic
oscillator potential
\begin{equation}
V(\mbox{\boldmath $\varrho$})=m^*\omega_0^2\varrho^2/2,
\label{eq:2Dharmonic}
\end{equation}
where $m^*$ is the effective mass of the host semiconductor.
\begin{figure}[h]
\setlength{\unitlength}{1mm}
\begin{picture}(120,54)
\put(-19,-24){\epsfig{file=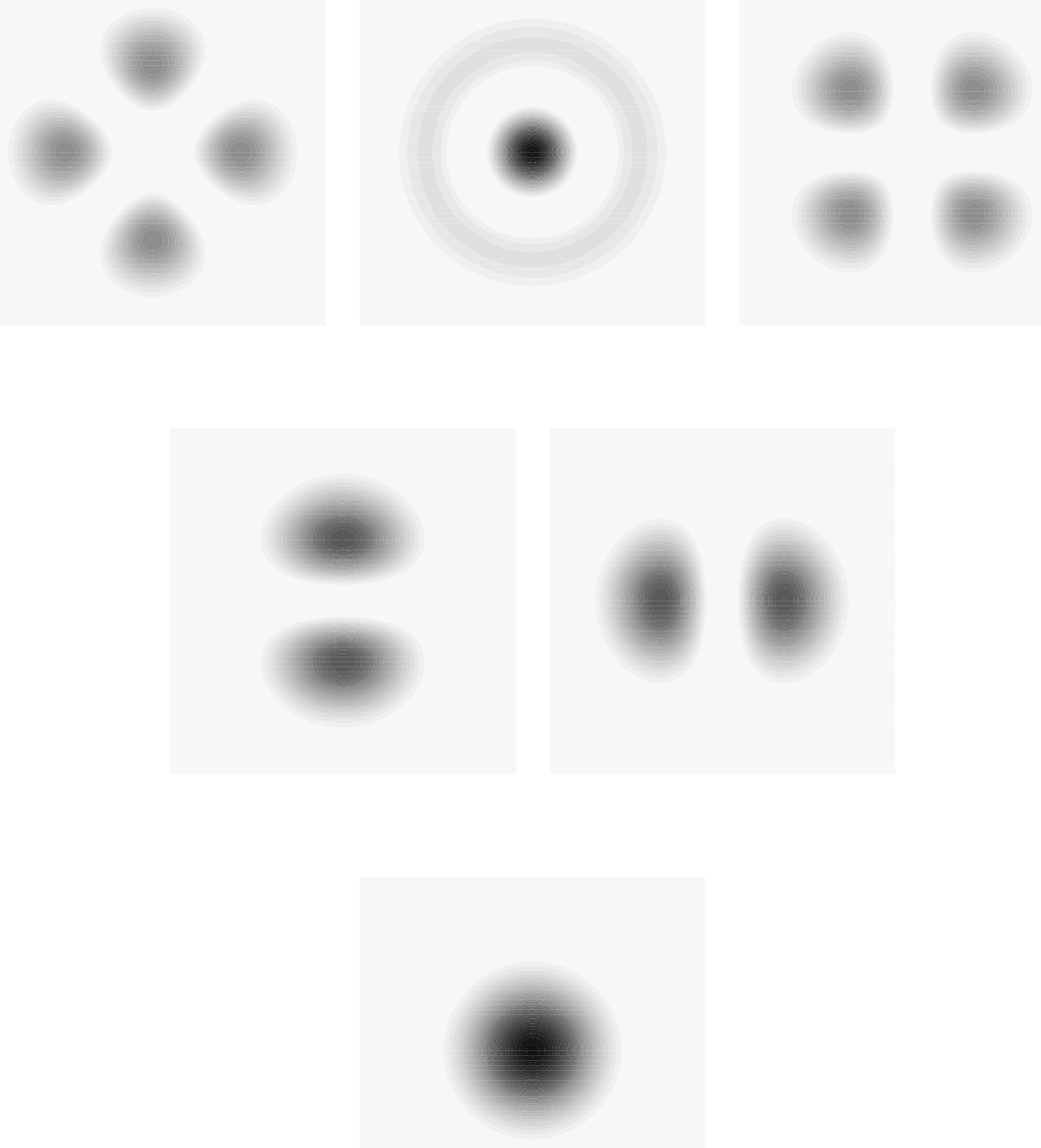,angle=0,width=7.3cm}}
\put(63,7){\epsfig{file=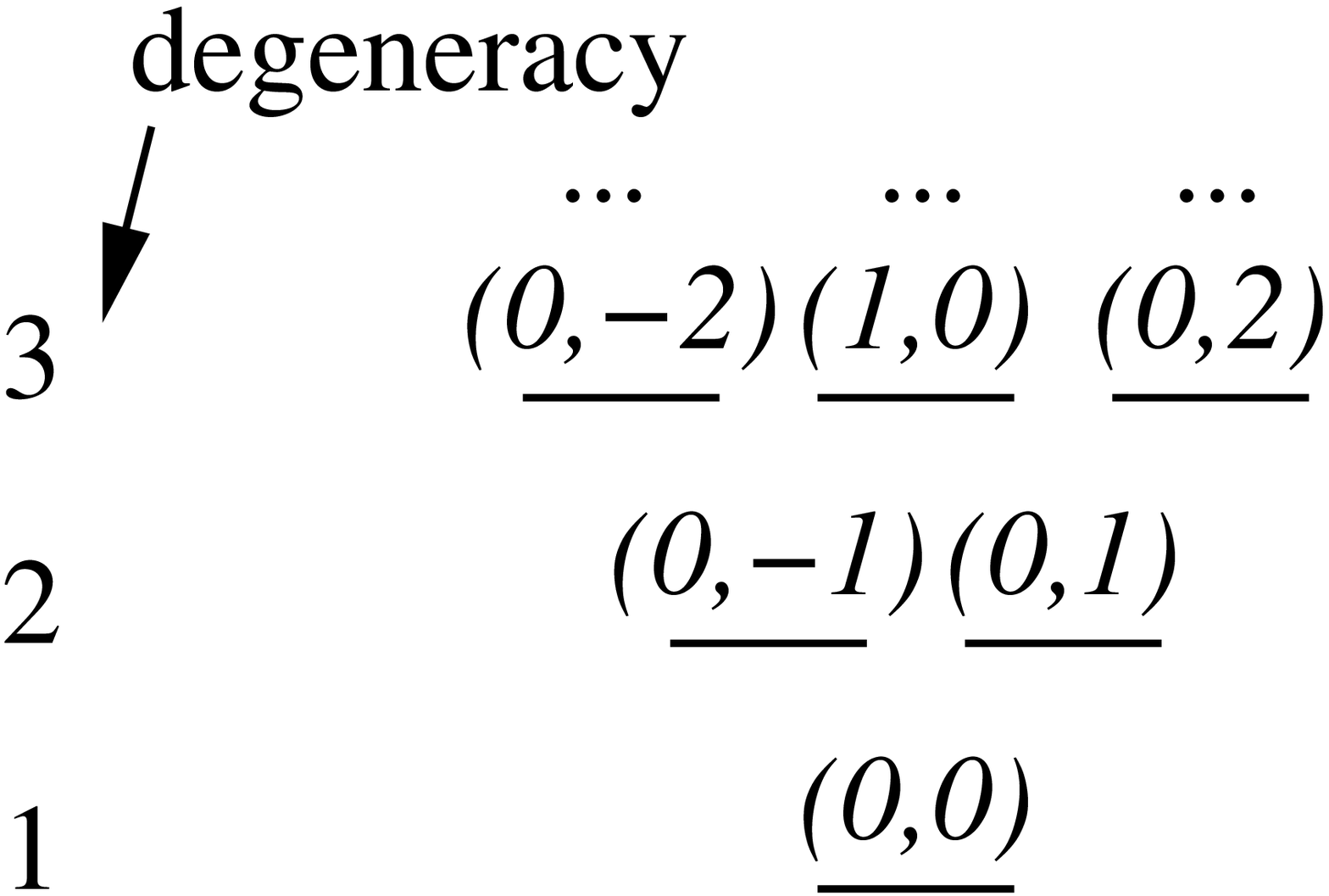,angle=-0,width=4.5cm}}
\put(112,3){\epsfig{file=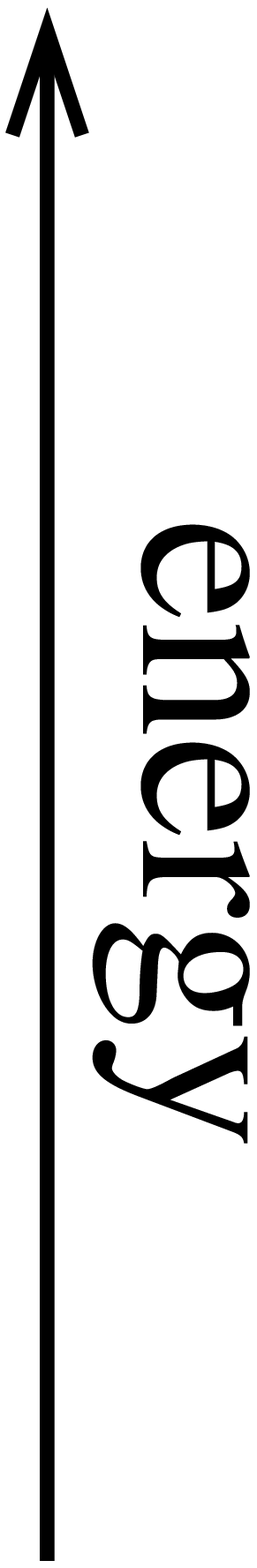,angle=0,width=0.6cm}}
\put(115,42){\Large (b)}
\put(47,42){\Large (a)}
\end{picture}
\caption{Two dimensional harmonic trap: The energy structure at zero field.
(a) Contour plot in the $x-y$ plane of the probability density of the
first lowest-energy single-particle orbitals.
(b) Energy shell structure.
The couple of numbers $(n,m)$ refer to the radial and azimuthal
single-particle quantum numbers, respectively.
}\label{fig2}
\end{figure}
Adequacy of Eq.\ (\ref{eq:2Dharmonic}) has been demonstrated both
by theoretical calculations \cite{Kumar} and far infra-red
spectroscopy \cite{QDreview}; in addition, it is the lowest order
approximation in the Taylor expansion of the weak electrostatic
potential.

The single-electron
Hamiltonian $H_0(\mbox{\boldmath $r$},s_z)$ is
\begin{equation}
H_0(\mbox{\boldmath $r$},s_z)=
(-\mathrm{i}\hbar\nabla+\left| e\right|\mbox{\boldmath $A$}/c)^2/2m^*+
V(\mbox{\boldmath $r$})+g^*\mu_{\mathrm{B}}Bs_z,
\label{eq:singleparticle}
\end{equation}
where we included a magnetic field parallel to the $z$ axis
$\mbox{\boldmath $B$}=B\hat{z}$
which couples with both the spin degree of freedom, $s_z=\pm 1/2$,
and the orbital motion via the vector potential $\mbox{\boldmath $A$}
=\mbox{\boldmath $B$}\times\mbox{\boldmath $\varrho$}/2$.
Note that, since the Bohr magneton of the {\em free} electron
$\mu_{\mathrm{B}}$ enters (\ref{eq:singleparticle}),
the spin part is usually negligible with respect to the orbital part
($g^*$ is the effective giromagnetic factor).

\begin{figure}[h]
\setlength{\unitlength}{1mm}
\begin{picture}(120,85)
\put(5,0){\epsfig{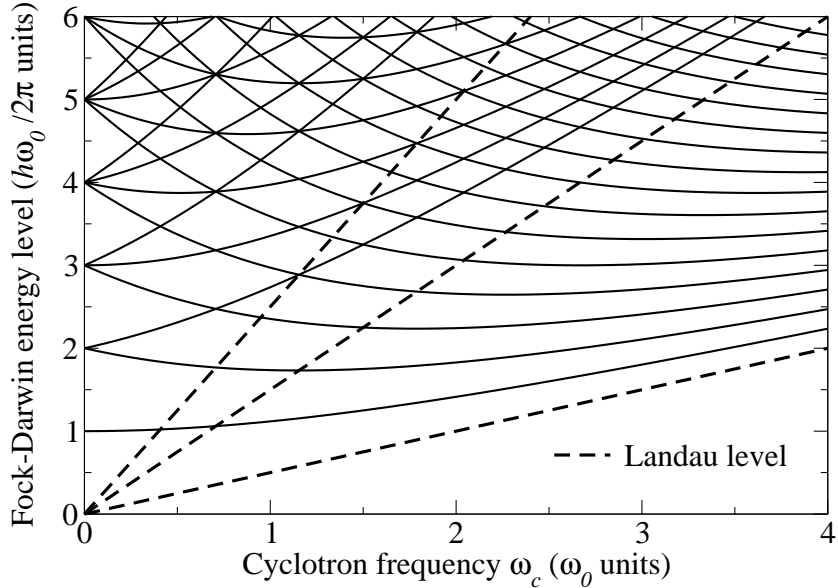}}
\end{picture}
\caption{Two dimensional harmonic trap: The single-particle Fock-Darwin levels
in a magnetic field $B$. The energies are given in units of $\hbar\omega_0$.
The field dependence is contained in the cyclotron frequency $\omega_c^*=
eB/m^*c$. The first three lowest-energy Landau levels are also depicted.
}\label{fig3}
\end{figure}

Because of the decoupling (\ref{eq:decoupling}), we can write the
eigenfunctions of the orbital part of the Hamiltonian
(\ref{eq:singleparticle}) as \protect{$\psi_{nm}(\mbox{\boldmath
$r$})=\varphi_{nm} (\mbox{\boldmath $\varrho$})\, \chi(z)$}, where
$\chi(z)$ is the ground state for the motion along $z$ and
$\varphi_{nm}(\mbox{\boldmath $\varrho$})$ are the eigenstates of
the 2D harmonic oscillator (Fock-Darwin levels, Fig.~\ref{fig2}
and \ref{fig3}).
The eigenvalues $\varepsilon_{nm}$ are given by
\begin{equation}
\varepsilon_{nm}=\hbar\Omega ( 2n + \left|m\right| + 1) - \hbar\omega^*_c
m/2,
\label{eq:eigenvalue}
\end{equation}
where $n$ ($n=0,1,2,\ldots$) and $m$ ($m=0,\pm 1,\pm 2,\ldots$) are
the radial and azimuthal quantum numbers, respectively, $\omega_c^*$
is the cyclotron frequency ($\omega_c^*=eB/m^*c$),
and \protect{$\Omega=(\omega_0^2 +\omega_c^{*2}/4)^{1/2}$} \cite{QDreview}.
When $B=0$ ($\omega_c^*=0$) the Fock-Darwin spectrum shows
a characteristic shell structure (Fig.~\ref{fig2}):
the level degeneracy linearly increases with the shell number
[Fig.~\ref{fig2}(b)],
and the energy spacing between neighboring shells is constant (see
Fig.~\ref{fig2}).

\begin{figure}[h]
\setlength{\unitlength}{1mm}
\begin{picture}(120,90)
\put(0,70){\epsfig{file=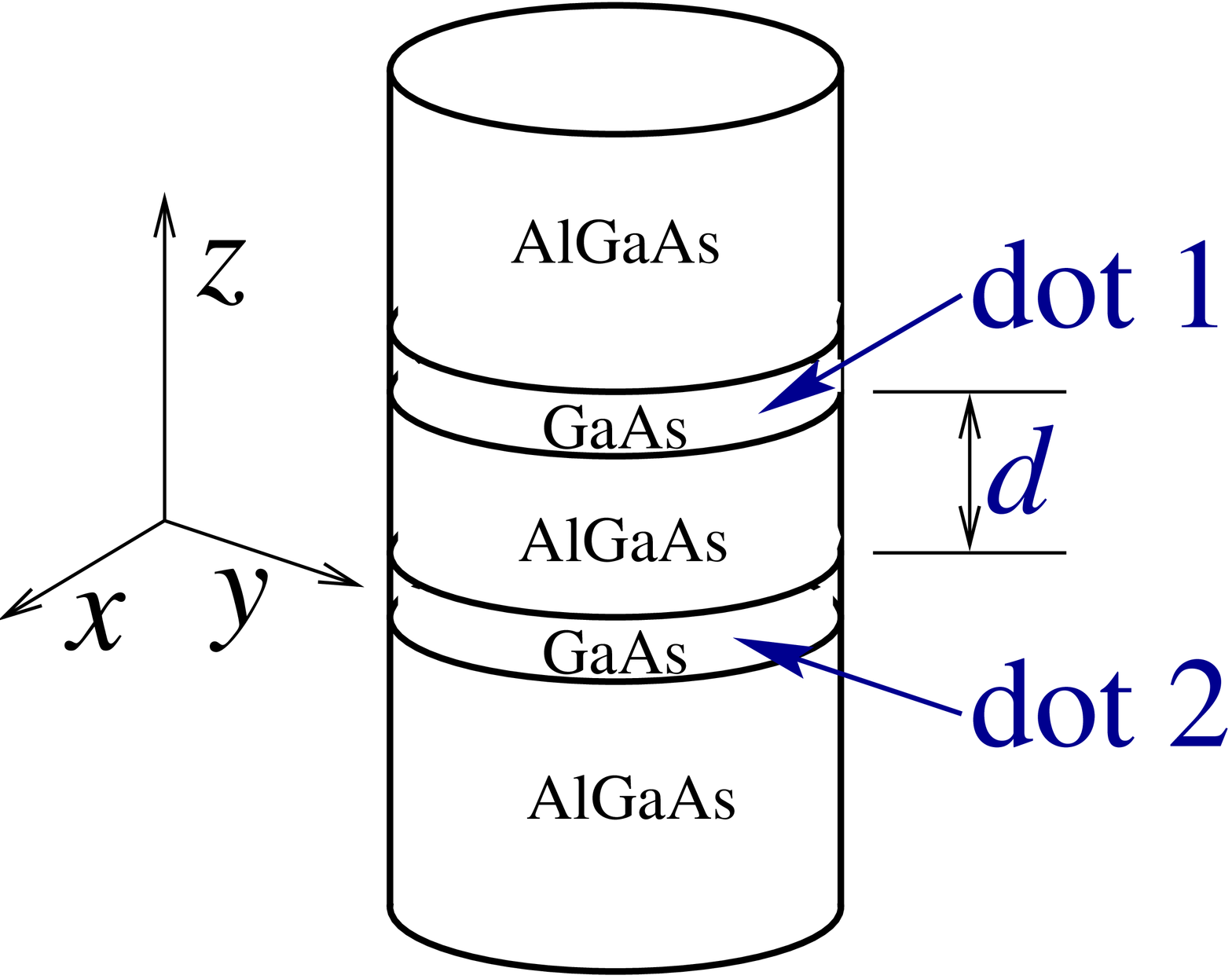,angle=-90,width=5.0cm}}
\put(5,66){\epsfig{file=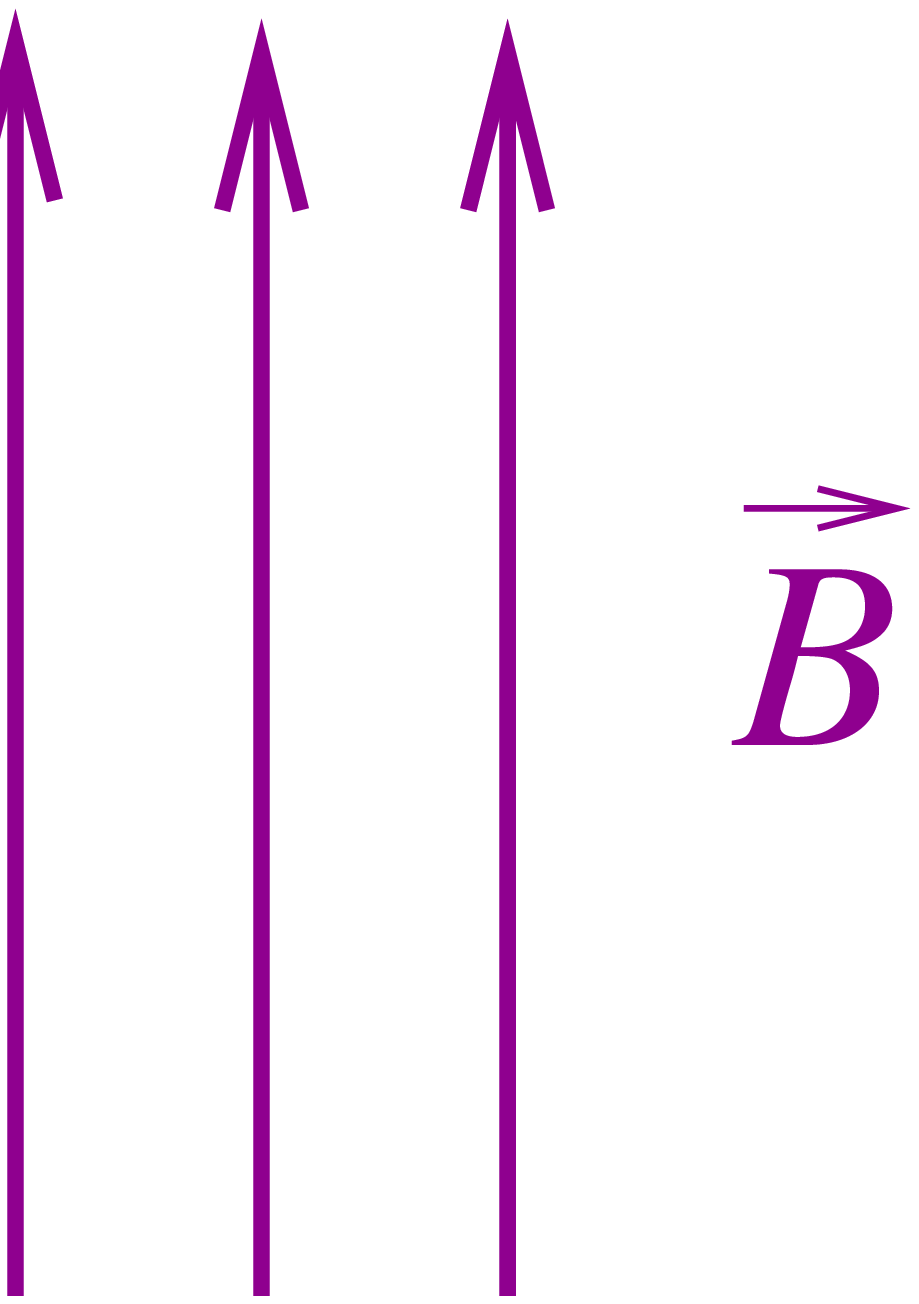,angle=0,width=1.3cm}}
\put(53,47){\epsfig{file=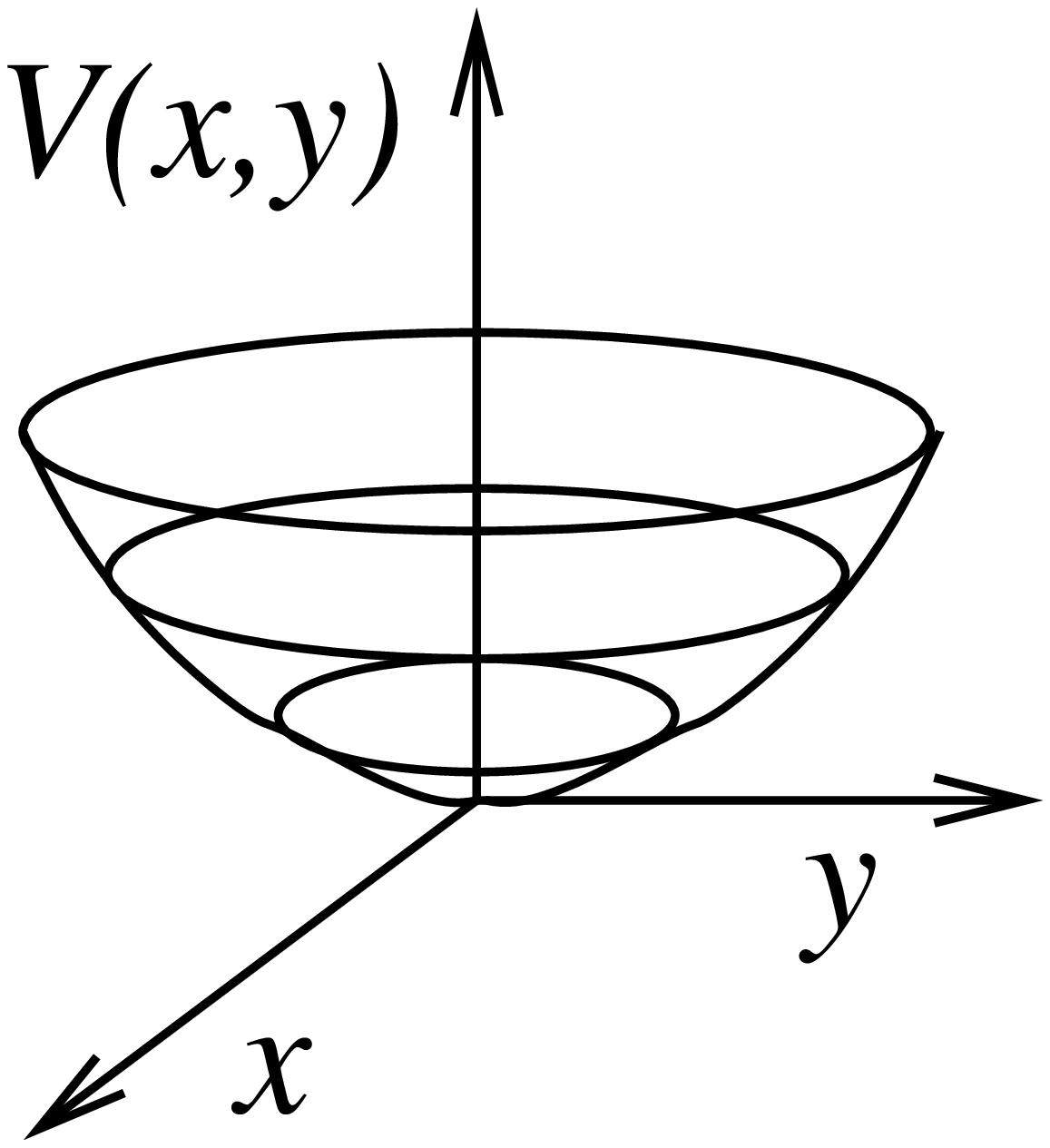,angle=0,width=3.2cm}}
\put(91,41){\epsfig{file=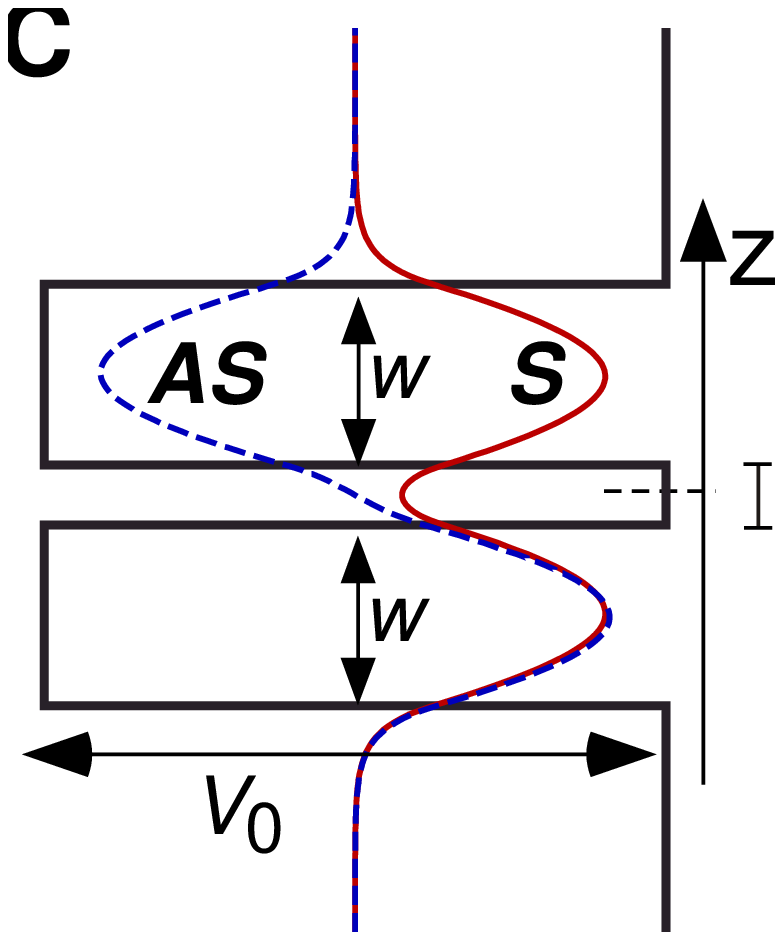,angle=0,width=3.2cm}}
\put(70,2){\epsfig{file=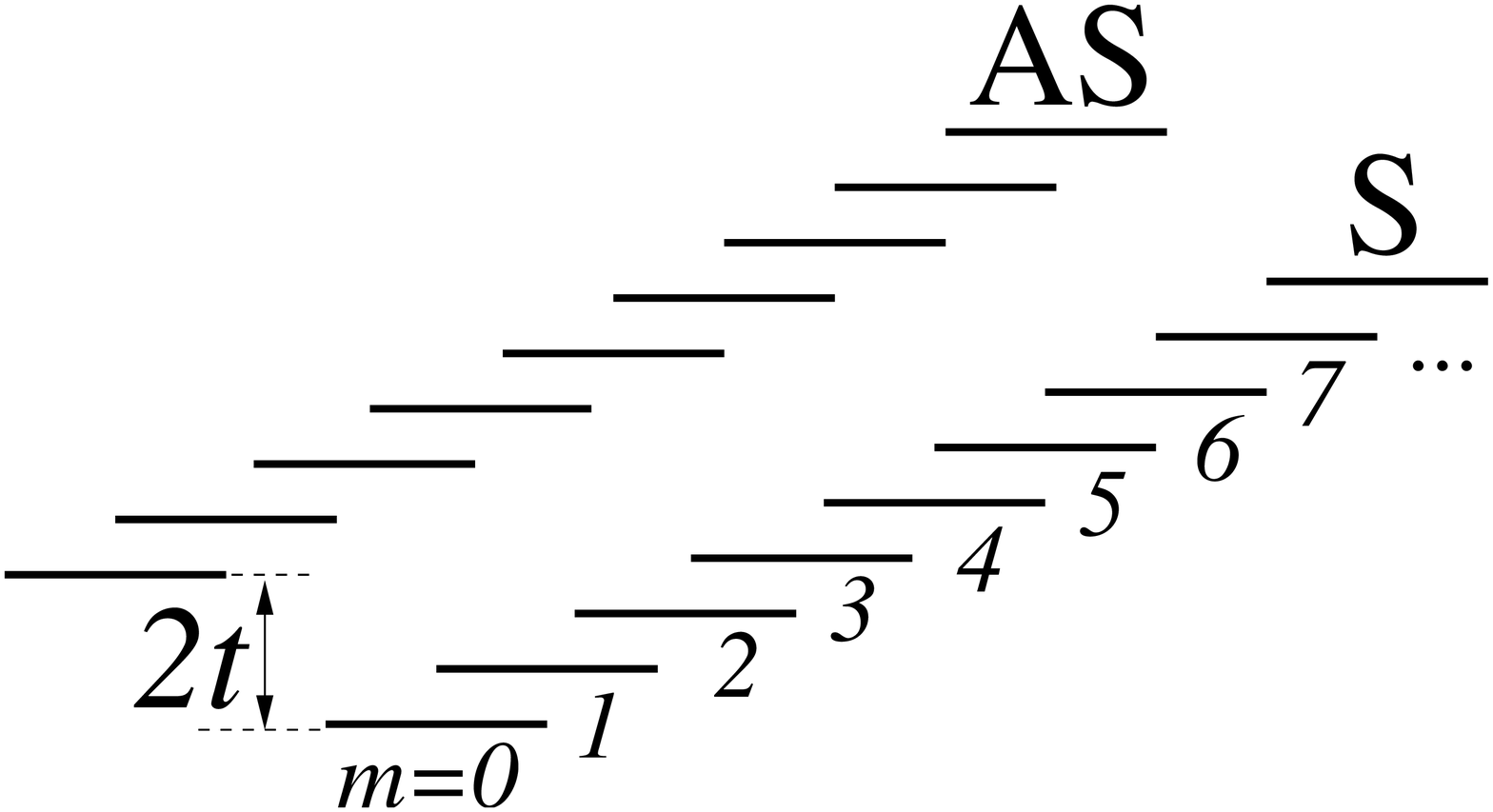,angle=0,width=5.5cm}}
\put(5,2){\epsfig{file=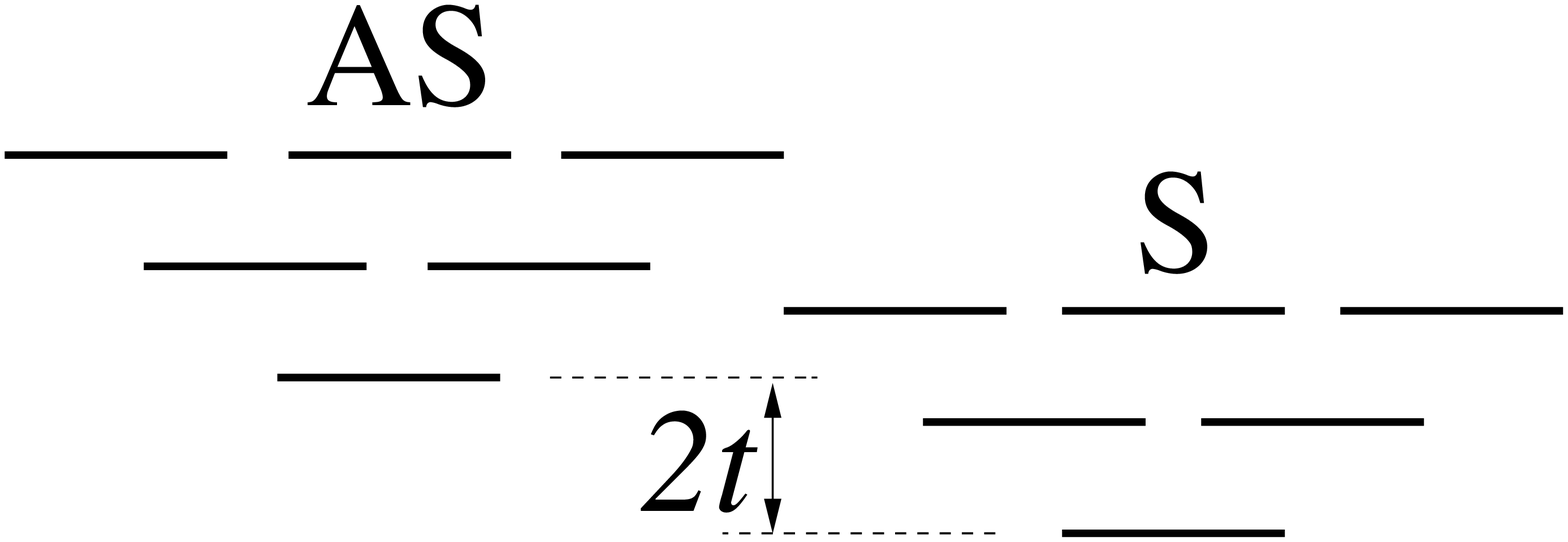,angle=0,width=5.5cm}}
\put(124,60){\Large $d$}
\put(49,20){\Large (d)}
\put(27,80){\Large (a)}
\put(70,20){\Large (e)}
\put(74,80){\Large (b)}
\put(108,80){\Large (c)}
\put(89,78){\epsfig{file=empty.eps,angle=0,width=0.7cm}}
\end{picture}
\caption{Model of the vertical artificial molecule. (a) Sketch of the
model device we consider. (b) In-plane single-particle
confinement potential. (c) Confinement potential along the $z$-axis.
$V_0$ and $d$ are respectively the height and the width of the
inter-dot potential barrier, while $w$ is the thickness of each
of the two dots.
(d) The symmetric (S) and antisymmetric (AS) sets of Fock-Darwin levels
at zero field. $t$ is the tunneling energy.
(e) The S and AS sets of Fock-Darwin levels in a high magnetic field.
Only the first lowest-energy states with $n=0$ and $m>0$ are depicted.
}\label{fig4}
\end{figure}

A magnetic field parallel to the growth direction splits the
degeneracies of the states (Fig.\ \ref{fig3}) and ``squeezes'' the
orbitals, as it appears from the expression of the characteristic
length $\ell=(\hbar/m^*\Omega)^{1/2}$, corresponding to the
average value of $\varrho$ on the ground state $\varphi_{00}$: the
stronger the field, the smaller the radius of the wavefunction. At
intermediate fields ($\omega_c\sim\omega_0$) the energy levels may
increase or decrease as a function of the field, depending whether
the azimuthal quantum number $m$ is negative or positive; at large
fields, however, all levels tend to the 2D highly degenerate
Landau levels. The latter can be obtained by letting
$\omega_c^*/\omega_0\rightarrow \infty$ in
Eq.~(\ref{eq:eigenvalue}). Accordingly, at large fields only
positive $m$ states are occupied ($n=0$). In this limit, the
energy separation $\Delta \varepsilon$ between levels differing on
$\left|\Delta m\right|=1$ becomes
\protect{$\Delta\varepsilon\approx \hbar\omega_0^2/\omega_c^*$},
which is $\propto 1/B$.

Additional flexibility in controlling the energy spectrum is given
by the possibility to grow QDs coupled by quantum mechanical
tunnelling. A typical ``vertical'' device is sketched in
Fig.~\ref{fig4}(a) (see ref.~\cite{Guy} for an experimental
realization). In this case, the carrier dynamics is not strictly
2D anymore, as two levels $\chi_i(z)$ [symmetric ($i=\mathrm{S}$)
and anti-symmetric ($i=\mathrm{AS}$) for a ``homonuclear''
molecule, see Fig.~\ref{fig4}(c)] enter the relevant energy range.
Correspondingly, the single particle spectrum comprises two sets
of Fock-Darwin states at small [Fig.~\ref{fig4}(d)] or high
[Fig.~\ref{fig4}(e)] magnetic fields. The energy separation
between S and AS states is $2t$, $t$ being the tunneling energy
which can be controlled, e.g., by varying the width $d$ or the
height $V_0$ of the inter-dot potential barrier while growing the
sample [Fig.~\ref{fig4}(c)].

When we consider the Coulomb interaction between carriers,
no exact solutions are available except very special
cases for $N=2$ \cite{Taut}. The interacting Hamiltonian
\begin{equation}
{\cal H}=\sum_{i=1}^N H_0(\mbox{\boldmath $r$}_i,s_{zi})+
\frac{1}{2}\sum_{i\neq j}\frac{e^2}{\kappa_{\mathrm{r}}
\left|\mbox{\boldmath $r$}_i-\mbox{\boldmath $r$}_j\right|},
\label{eq:Htot}
\end{equation}
with $\kappa_{\mathrm{r}}$ static dielectric constant,
must be solved numerically. Dramatic alterations of the
few-body energy spectrum and wavefunctions appear,
depending on the relative ratio of the magnitudes of the
one- and two-body terms in ${\cal H}$.

If the single-electron term $H_0$ prevails, the
system basically behaves as a non-interacting system, or,
better, as a Fermi liquid with corrections due to the residual
part of the interaction. This is the regime where the periodic
table of artificial atoms has been observed \cite{PRLTarucha}.
If instead the Coulomb term dominates, electrons undergo
a transition to a qualitatively new state where they orderly
arrange themselves in space in such a manner to minimize
the electrostatic energy \cite{QDreview,MaksymPRL,Bryant}.
In this few-body strongly correlated
regime, which is reminescent of the Wigner crystal phases in
extended systems \cite{Wigner}, the physics turns out to be classical.
Asymptotically, all operators become commutative and all spin configurations
tend to perfect degeneracy (except possibly the Zeeman splitting).

The Wigner regime can be artificially driven in two ways. One
possibility is fabrication of QDs with weak enough lateral
confinement that the average electron density $n$ is extremely
small: since the 2D one-electron term in Eq.~(\ref{eq:Htot}) goes
like $n$ while the two-body term like $n^{1/2}$, in the dilute
limit the former becomes negligible with respect to the latter
\cite{Bryant}. Another possibility is to apply a magnetic field
strong enough that the energy separation $\Delta\varepsilon$
becomes small compared to the typical Coulomb energy
(equivalently, $\ell\ll n^{-1/2}$): again, the interaction term of
${\cal H}$ controls the low-energy physics. We focus on this case
in Sec.~\ref{Wigmol}.

\begin{figure}[h]
\centerline{\epsfig{file=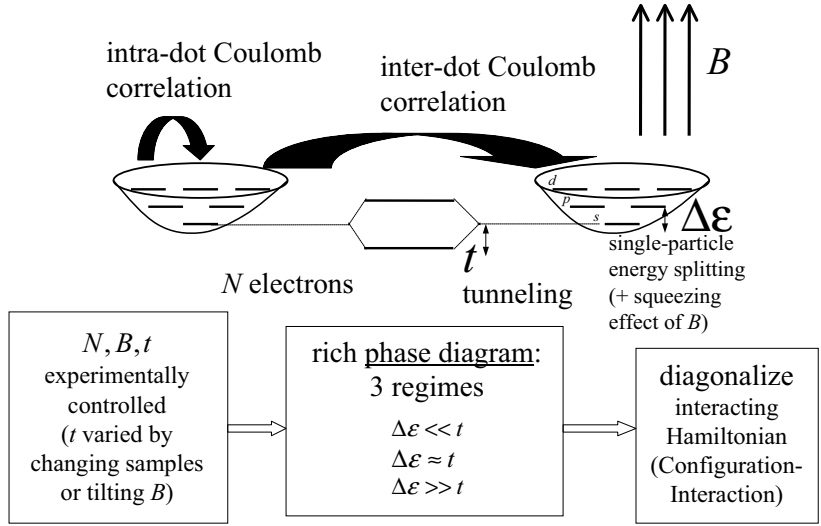,angle=90,width=11.6cm}}
\caption{Artificial molecules: The relevant energy scales
controlling the few-electron interacting ground state.
Many of these energies can be artificially tuned together
with the number of electrons $N$.
}\label{fig5}
\end{figure}
Figure \ref{fig5} summarizes the relevant energy scales for a
coupled QD system and how they can be tailored. Both single
particle energies and tunneling can be independently tuned by
device engineering or, more practically, by external magnetic
field intensity and direction. The ``degree'' of correlation
of the system can be similarly controlled, as well as $N$.

\section{Structural transitions in Wigner molecules}\label{Wigmol}

We now discuss the ground state of $N$ electrons in coupled QDs in
a high magnetic field $B$, i.e., such that single-QD correlation
functions show strong localization \cite{us,epl}. We present
results for $N=6$ \cite{epl}. The system with $N<6$ exhibits a
similar physics. We can identify three regimes corresponding to
different electron arrangements: (I) At small $d$, tunneling
dominates and the system behaves as a unique coherent system. (II)
As $d$ is increased, all energy scales become comparable. (III)
When eventually tunneling is suppressed, only the ratio between
intra- and inter-dot interaction is the relevant parameter for the
now well separated QDs.

\begin{figure}[h]
\setlength{\unitlength}{1mm}
\begin{picture}(120,125)
\put(0,0){\epsfig{file=angular-NATO.eps,angle=0,width=8.5cm}}
\put(93,92){\epsfig{file=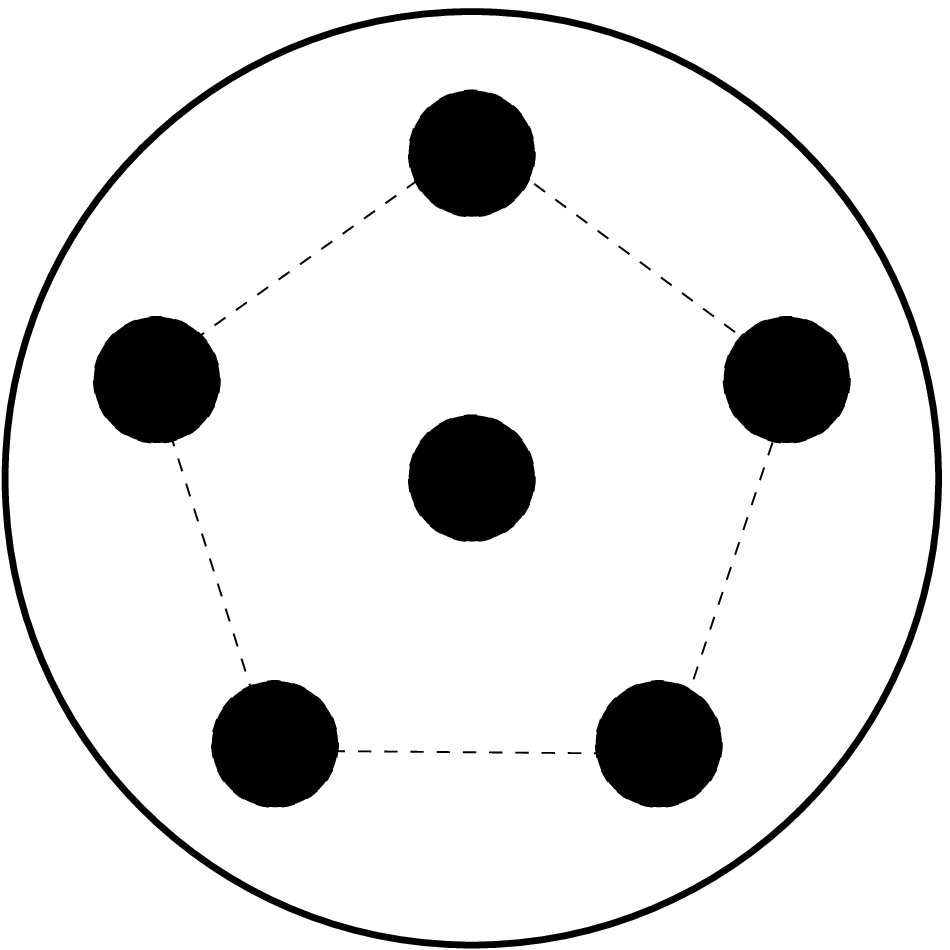,angle=0,width=2.5cm}}
\put(95,87){\LARGE $\Delta\varepsilon\ll t$}
\put(93,55){\epsfig{file=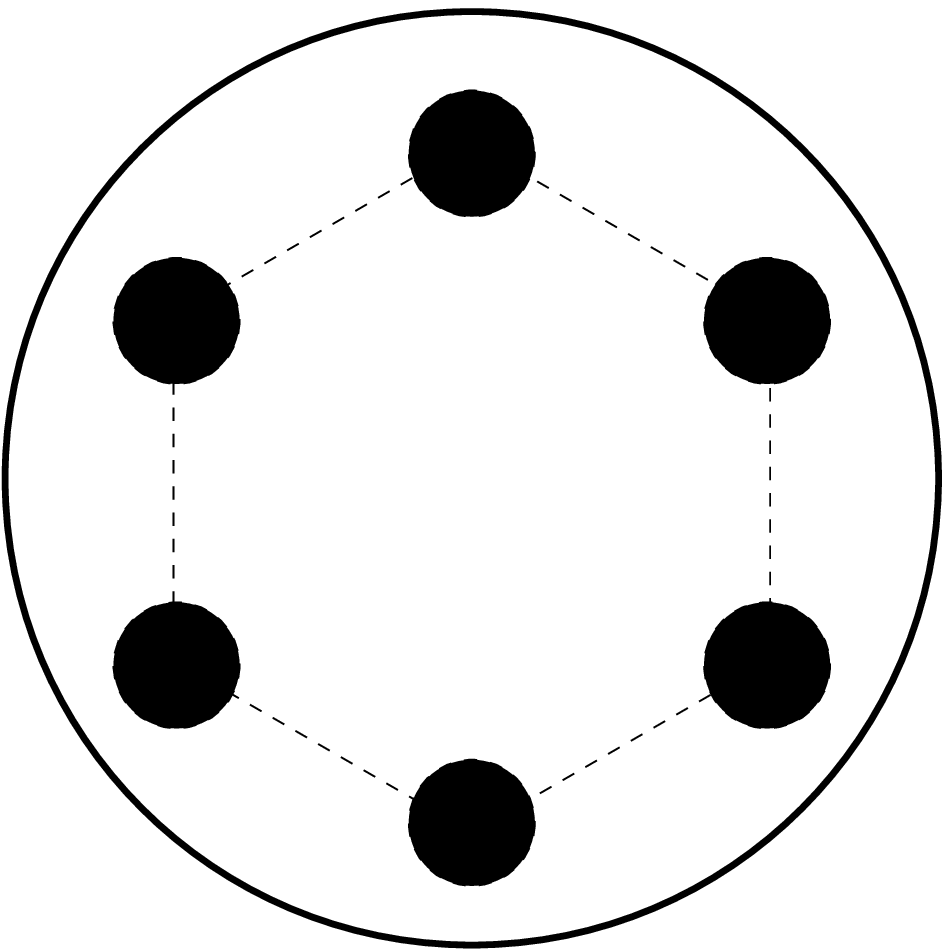,angle=0,width=2.5cm}}
\put(95,50){\LARGE $\Delta\varepsilon\approx t$}
\put(93,18){\epsfig{file=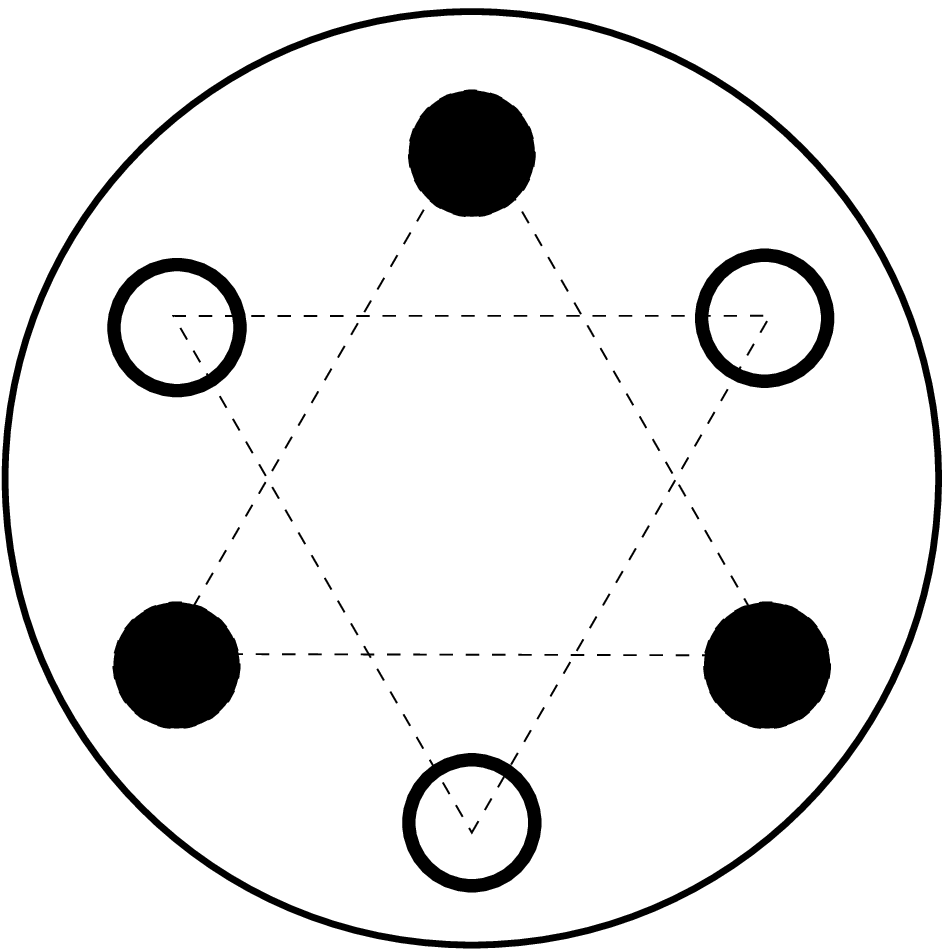,angle=0,width=2.5cm}}
\put(95,13){\LARGE $\Delta\varepsilon\gg t$}
\put(50,89){\epsfig{file=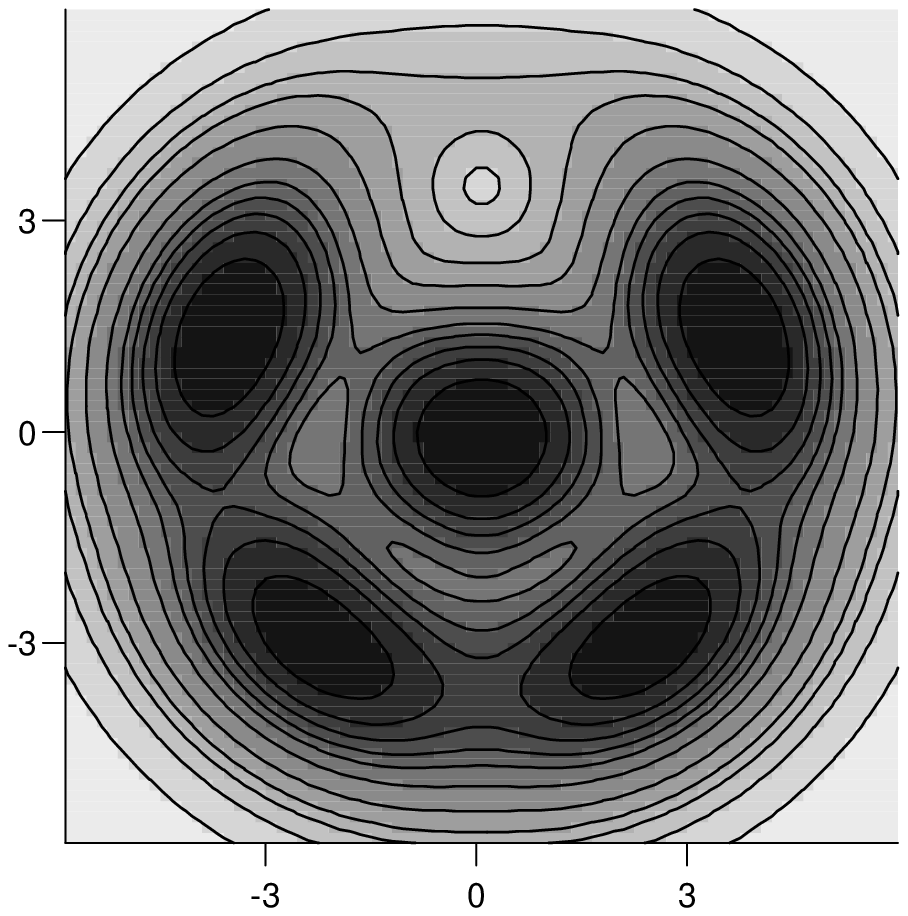,angle=0,width=4.0cm}}
\put(77,91){$x$}
\put(53,115){$y$}
\put(59.5,96.5){\epsfig{file=tondo.eps,angle=0,width=1.6cm}}
\end{picture}
\caption{The three phases of the Wigner molecule as
the critical ratio $\Delta\varepsilon/t$ is changed.
Here we numerically solve the Hamiltonian of Eq.~(\ref{eq:Htot})
by means of the Configuration Interaction method, namely
we expand the few-body wavefunction as a linear combination
of Slater determinants made by filling with $N$ electrons
the Fock-Darwin orbitals. We consider only levels with
$m>0$ and $n=0$. Further details in ref.~\cite{epl}.
$B$ is fixed (25 T) and we assume as parameters (see text and
caption of Fig.~\ref{fig4})
$\hbar\omega_0$=3.70 meV, $w$=12 nm, $V_0$=250 meV,
$m^*$=0.067$m_e$, $\kappa_{\mathrm{r}}$=12.4, $g^*$=-0.44
as typical values of realistic devices \cite{Guy}.
The three phases correspond to $d$=2, 4.6, 8 nm, respectively.
}\label{fig6}
\end{figure}
To analyze the ground state of the artificial molecule in the
different regimes, we show in Fig.~\ref{fig6} the pair
correlation function $P(\mbox{\boldmath $\varrho$} , z ;
\mbox{\boldmath $\varrho$}_0 ,
z_0 ) =\sum_{i\neq j}\left<\delta
(\mbox{\boldmath $\varrho$}-
\mbox{\boldmath $\varrho$}_i) \delta ( z - z_i ) \delta
(\mbox{\boldmath $\varrho$}_0-
\mbox{\boldmath $\varrho$}_j) \delta (z_0 - z_j)
\right>/N(N-1)$ (the average is on the ground state)
\cite{us,epl}. Since we are interested in the strong localization
regime, we expect that spin texture does not alter the
essential physics, therefore we assume that electrons are
spin polarized \cite{Carlos}.
Figure \ref{fig6} shows $P( \mbox{\boldmath $\varrho$} , z ;
\mbox{\boldmath $\varrho$}_0 , z_0
)$ along a circle in the same dot (solid line) or in the opposite
dot (dashed line) with respect to the position of a reference
electron, taken at the maximum of its charge density,
$(\mbox{\boldmath $\varrho$}_0 , z_0)$. The right column shows the electron
arrangement in the QDs as inferred by the maxima of $P(
\mbox{\boldmath $\varrho$} , z ; \mbox{\boldmath $\varrho$}_0 , z_0 )$.

At small $d$ the whole system is coherent, i.e., it behaves as a
unique QD. The electrons, delocalized over the dots, arrange at
the vertices and the center of a regular pentagon (Phase I). At
intermediate values of the tunnelling energy electrons sit at the
vertices of a regular hexagon. Contrary to the previous case peaks
in the upper and lower dots have different heights (Phase II).
Finally, when $d$ is sufficiently large, the structure evolves
into two isolated dots coupled only via Coulomb interaction.
Accordingly, three electrons in each dot sit at the vertices of
two equilateral triangles rotated by 60 degrees (Phase III).

It is important to note from Fig.~\ref{fig6} that Phase I and III
are strongly localized phases, where quantum fluctuations play a
minor role, and electron configurations are basically determined
by Coulomb interactions; accordingly, they have completely
classical counterparts \cite{Peeters_classical}. On the contrary,
in Phase II tunneling fluctuations prevents electron from
localizing and therefore the configuration has a ``liquid''
character. Such phase cannot be explained in term of Coulomb
interactions solely and, in fact, the exagonal arrangement shown
in Fig.~\ref{fig6} is classically unstable.

This example shows that in artificial molecules at high $B$ one
can drive qualitative changes in the ground state of the
interacting electrons that are clearly precursors of quantum phase
transitions in the infinite system \cite{QPT}. While the
observation of such transitions in the bulk is very difficult, QDs
seem to constitute ideal systems to explore in the laboratory the
fundamentals of electron correlation. For example, it is
conceivable that inelastic light scattering spectroscopy is able
to probe the different roto-vibrational ``normal modes'' of the
three phases, thus allowing for their experimental detection. See
ref.~\cite{epl} for further details.


\begin{acknowledgements}
This work was supported by INFM (SSQI), by MIUR (FIRB Quantum Phases
of Ultra-Low Electron Density Semiconductor Heterostructures),
by MAE (Progetto di Particolare Rilevanza ``Controllo di stati di carica
e spin in punti quantici''), and by the EC (SQID).
\end{acknowledgements}

\index{first}
\end{article}


\end{document}